\documentclass[10pt, conference]{IEEEtran}
\IEEEoverridecommandlockouts
\usepackage{cite}
\usepackage{amsmath,amssymb,amsfonts}
\usepackage{algorithmic}
\usepackage{graphicx}
\usepackage{textcomp}
\usepackage{xcolor}
\usepackage{caption}
\usepackage{subcaption}

\def\BibTeX{{\rm B\kern-.05em{\sc i\kern-.025em b}\kern-.08em
    T\kern-.1667em\lower.7ex\hbox{E}\kern-.125emX}}

\usepackage{url}

\begin{document}

\title{Quantum Simulations for Carbon Capture on Metal-Organic Frameworks}

\author{\IEEEauthorblockN{Gopal Ramesh Dahale}
\IEEEauthorblockA{\textit{Alumni of Indian Institute of Technology Bhilai, India}\\
Navi Mumbai, India \\
dahalegopal27@gmail.com}
}

\maketitle

\begin{abstract}
Direct air capture of Carbon Dioxide is a technical solution that does not rely on natural processes to capture CO$_2$ from the atmosphere. In DAC, the filter material is designed to specifically bind CO$_2$ molecules. Hence a high-capacity filter is sought. We aim to leverage the potential of quantum computing to improve the filters used in DAC. Metal-Organic Frameworks (MOFs) have high surface area and tunable pore sizes which makes them an attractive material for gas storage and separation. Using the variational quantum eigensolver (VQE) algorithm, we find the minimum of the potential energy surface (PES) by first considering only the active site of the MOF (the metal ion). For complex systems, we employ Density Matrix Embedding Theory and use VQE as a fragment solver at the binding site. Techniques like deparameterisation are used to minimise the count of trainable parameters. We present results of ideal and noisy simulations as well as from a real hardware device. Resources are estimated for MOFs unit cell. The findings from our study demonstrates the potential of quantum computing to effectively perform quantum simulations of strongly correlated fragments.
\end{abstract}


\section{Introduction}
Carbon dioxide emissions are one of the main contributors to global warming. More than 410 parts per million of CO$_2$ are now present in the earth's atmosphere, up from a pre-industrial level of 280 parts per million. In prehistoric times, it would take between 5000 and 20,000 years for CO$_2$ levels to rise to the same level as they have in the past 60 years as a result of human activity \cite{carbonlevel}. The entire earth system is then severely impacted by this. Therefore, it is crucial to significantly reduce manmade CO$_2$ emissions to halt climate change. 

Direct air capture (DAC) is a technological approach to removing CO$_2$ from the atmosphere that does not rely on organic processes. In DAC, an air intake fan is positioned in front of a designed CO$_2$ filter. The filter material is intended to bind CO$_2$ molecules only, ideally without capturing any additional atmospheric gas molecules that would cause the filter to clog. The technology's energy demand is a major challenge. Due to the dilution of CO$_2$ in the Earth's atmosphere, a substantial volume of air must pass through the filter for effective CO$_2$ capture. The operation of fans and the regeneration of filters require significant energy. The capacity of the filter, determining the ratio of capture time to release time during the two phases of CO$_2$ collection and release, is directly linked to its performance. Hence, filters with high CO$_2$ adsorption capacity are preferred. The solution to this difficulty lies in utilising the capability of quantum computers to enhance the filters employed in the DAC of atmospheric CO$_2$.

Over the years, several materials and technologies have been created for the adsorption of CO$_2$; these include metal oxides, zeolites, activated carbons, fluorinated solvents, molecular sieves, and metal-organic frameworks (MOFs).  MOFs are in demand for carbon capture applications due to their typically high surface area, variable pore size, and low heat capacity. Their structures must be specially created to guarantee excellent CO$_2$ uptake capability and CO$_2$ selectivity. They contain open metal sites (OMSs), which can serve as CO$_2$ adsorption sites \cite{mof}. However, it is worth noting that other materials, such as amines are also well-known for their reversible reactions with CO$_2$, which make them ideal for the separation of CO$_2$ from many CO$_2$-containing gases, including flue gas. Techniques like amine scrubbing are used for post-combustion CO$_2$ capture \cite{aminereview}.

One of the frontiers for the Noisy Intermediate-Scale Quantum (NISQ) era’s practical uses of quantum computers is quantum chemistry. By employing Hybrid Quantum Classical Optimisation we aim to investigate the minimum of the Ground Potential Energy Surface (PES) of the MOF with gas molecules with increasing the complexities of the system. We used a deparameterisation approach \cite{deparam} which helped in simplifying the energy landscape while maintaining the accuracy of the global minimum.

\section{Methods}

\subsection{Variational Quantum Eigensolver}
Variational Quantum Eigensolver (VQE) \cite{vqe} is one of the successful quantum algorithms for calculating the ground state energy of molecules. The ground state energy computation is viewed by the VQE method as an optimisation problem. The molecular Hamiltonian and a parameterized quantum circuit that prepares the molecule's quantum state are inputs, and the cost function to be minimised is determined by the Hamiltonian's expectation value. Therefore, finding a particular Hamiltonian's ground state is similar to optimising the circuit. In VQE, a classical optimizer performs the optimisation while a quantum computer assesses the cost function.

The choice of the ansatz used to approximate the ground state is important in determining a close upper bound to the actual energy. There exist problem-inspired ansatzes like UCCSD that utilize the excitations present within the electronic structure of the molecules. As we know, these problem-inspired ansatzes use multi-qubit gates and are high-depth circuits. These are not suitable for quantum hardware and are infeasible due to the limitations of NISQ devices. In this work, we focus on hardware-efficient ansatzes to improve the convergence of the VQE algorithm and make it feasible with current NISQ devices. We performed active space reduction by selecting the orbtials near HOMO-LUMO levels to reduce the orbitals count and therefore the number of qubits.

\subsection{Density Matrix Embedding Theory (DMET)}

Performing precise calculations of the electronic structure of large molecules typically come at a high computational expense. Consequently, it becomes crucial to use an effective technique for cutting the computing cost while keeping accuracy while performing electronic structure calculations when we focus on large-sized molecules, those vital for industrial problems. 

In DMET \cite{dmet1}\cite{dmet2} a molecule is broken down into fragments, and each fragment is viewed as an open quantum system that is entangled with every other piece, collectively constituting the fragment's environment (or ``bath"). According to this framework, a highly accurate quantum chemistry approach, such as the complete CI method or a coupled-cluster method, is used to solve the following Hamiltonian to determine the electronic structure of a given fragment.

\begin{eqnarray*}
     \sum_{rs}^{frag+bath}\left[ h_{rs} + \sum_{mn}[(rs|mn) - (rn|ms)]D_{mn}^{(mf)env}\right] a_r^{\dagger}a_s + \\
     \sum_{pqrs}^{frag+bath} (pq|rs) a_p^{\dagger}a_r^{\dagger}a_sa_q - \mu\sum_r^{frag}a_r^{\dagger}a_r
\end{eqnarray*}

The expression $\sum_{mn}[(rs|mn) - (rn|ms)]D_{mn}^{(mf)env}$ describes the quantum mechanical effects of the environment on the fragment, where $D_{mn}^{(mf)env}$  is the mean-field electronic density obtained by solving the Hartree–Fock equation. The Hamiltonian's one-particle term incorporates the environmental quantum mechanical effects. By adjusting $\mu$, the additional term $\mu\sum_r^{frag}a_r^{\dagger}a_r$ makes sure that the system's overall number of electrons and the number of electrons in each of the fragments are equal.

\subsection{Deparameterisation}\label{sec:deparam}

\cite{deparam} suggests that in some situations, when Hamiltonian is sufficiently sparse, producing a global minimum with exact energy does not need full parameterisation of all $R_y$ gates in an ansatz with depth $L_{min}$ (or greater). Thus, a heuristic deparameterisation approach can be used to further minimise the number of parameters. The easiest implementation is the one described below. 

A parameterised $R_y$ gate is initially chosen and frozen by giving it a set rotation value. The preferred rotation amplitude is zero, however other standardised values, such as $\pm \pi/2$ or $\pm\pi$, can also be utilised. The former option is preferred for several reasons, including 

\begin{itemize}
    \item The ability to transform the $R_y$ gate into a virtual identity gate that does not require actual implementation, which eliminates any related quantum gate noise.
    \item The computational cost of improving the parameter linked to that gate is likewise reduced in terms of software.
    \item Last but not least, we can build quantum circuits with more degrees of freedom for circuit translation by searching for as many virtual identity gates as possible. This may enable more effective mapping of the quantum circuit onto real quantum hardware and hence reduce its effective depth.
\end{itemize}

\section{Error Mitigation Techniques}

We used two main techniques: M3 \cite{m3} for noisy simulation and Twirled readout error extinction (T-REx) \cite{trex} for hardware experiments. Both have minimal mitigation costs and they mitigate readout errors. T-REx employs Pauli twirling as a method to mitigate noise arising from quantum measurements. The strength of T-REx lies in its ability to address noise without assuming any particular noise model, making it a versatile and highly efficient approach. On the other hand, M3 operates within a smaller subspace that is defined by the noisy input bitstrings requiring correction. This approach becomes advantageous because the number of distinct bitstrings can be significantly fewer than the dimensionality of the entire multi-qubit Hilbert space. As a result, the linear system of equations to be solved is considerably simplified, making the process more manageable and efficient.

\section{Results and Discussion}

\subsection{Metal ion and Gas Molecule}
\begin{figure}[h!]
\begin{minipage}[t]{0.475\columnwidth}
  \includegraphics[width=0.6\linewidth]{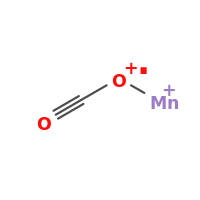}
    \caption{CO$_2$ + Mn (II) molecular system}
    \label{fig:co2_mn2_sys}
\end{minipage}\hfill 
\begin{minipage}[t]{0.475\columnwidth}
  \includegraphics[width=1.1\linewidth]{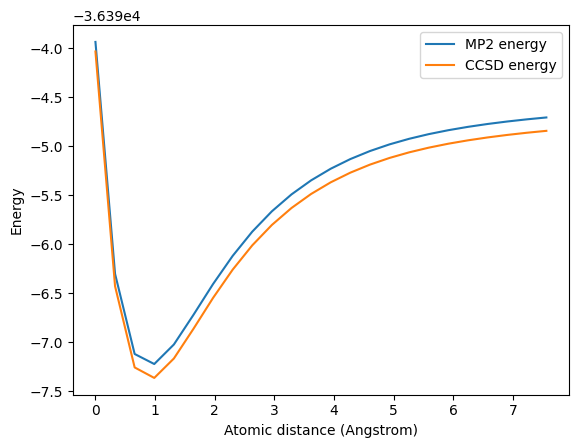}
    \caption{CCSD and MP2 PES for CO$_2$ + Mn (II) system}
    \label{fig:ref_co2_mn2}
\end{minipage}
\end{figure}

\begin{figure}[h!]
    \centering
    \includegraphics[width=0.7\linewidth]{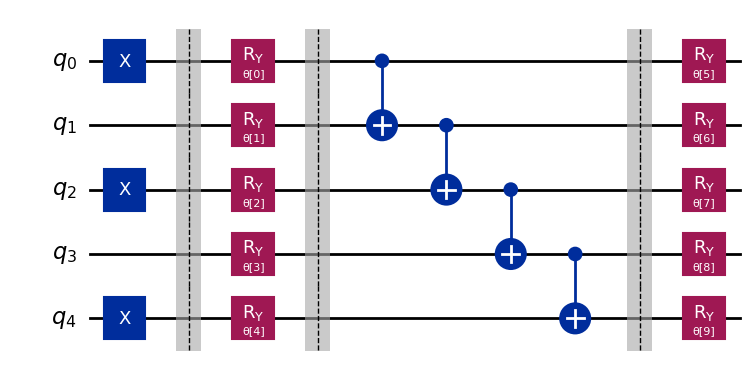}
    \caption{Hardware-efficient ansatz with 1 layer, 5 qubits and initial Hartree-Fock state}
    \label{fig:circ_co2_mn2}
\end{figure}

We investigated the minimum of the PES for CO$_2$ + Mn (II) molecular system shown in Fig.~\ref{fig:co2_mn2_sys}. Fig.~\ref{fig:ref_co2_mn2} shows the reference PES. For our analysis, we focus on the minimum of the PES. All calculations were performed using the cc-pVDZ basis. For a quantum simulation, we performed active space reduction to 4 spatial orbitals \cite{asr} which gives a quantum circuit of width 5 as shown in Fig.~\ref{fig:circ_co2_mn2}.

\subsubsection{Noiseless simulations}
\begin{figure}[h!]
    \centering
    \includegraphics[width=0.6\linewidth]{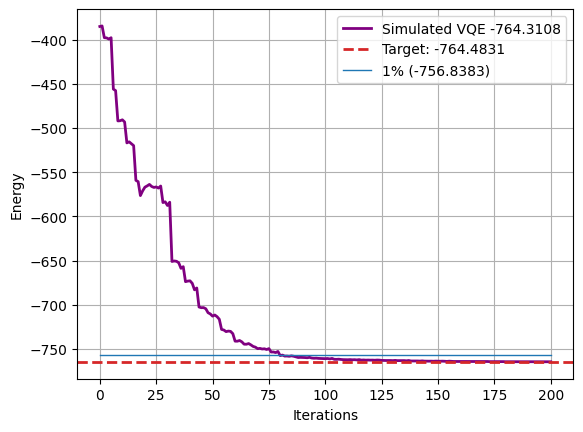}
    \caption{Convergence of VQE algorithm using SPSA optimizer}
    \label{fig:conv_co2_mn2}
\end{figure}

\begin{figure}[h!]
\begin{minipage}[t]{0.475\columnwidth}
  \includegraphics[width=0.9\linewidth]{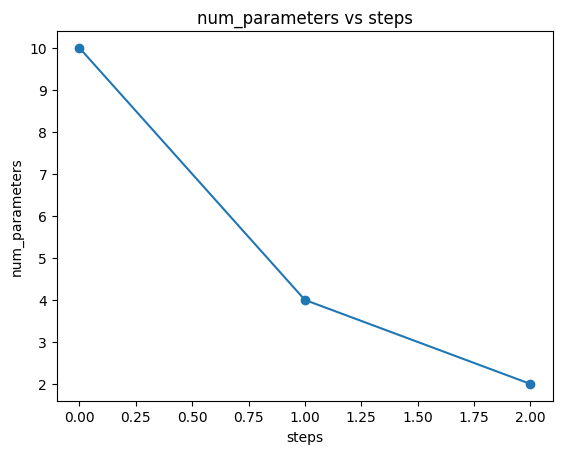}
  \caption{Reduction in trainable parameters after 3 steps of deparameterisation.}
  \label{fig:num_param}
\end{minipage}\hfill 
\begin{minipage}[t]{0.475\columnwidth}
  \includegraphics[width=\linewidth]{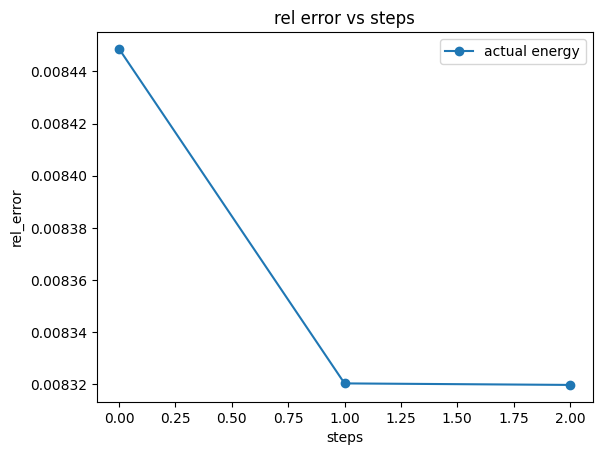}
  \caption{Relative error with deparamerisation steps.}
  \label{fig:rel_err}
\end{minipage}
\end{figure}

Parity fermion-to-qubit mapping \cite{parity} is used in all experiments unless stated. SPSA optimizer \cite{spsa} was used for optimization for 100 iterations. A successful convergence to the ground state is observed (Fig.~\ref{fig:conv_co2_mn2}). We performed the deparameterisation procedure as described in Section \ref{sec:deparam}. Fig.~\ref{fig:num_param} shows the reduction in the number of trainable parameters from 10 to 2 while retaining high accuracy for the global minimum Fig.~\ref{fig:rel_err}. The 8 gates were frozen to a value of $\pm \pi/2$.

\subsubsection{Noisy Simulations and Hardware executions}

\begin{figure}[h!]
    \centering
    \includegraphics[width=\linewidth]{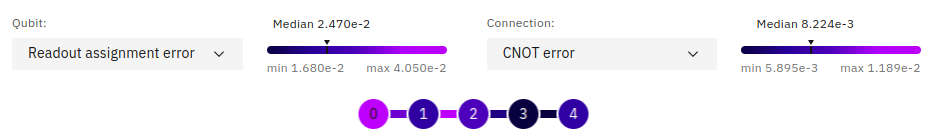}
    \caption{\texttt{ibmq manila} device with 5 qubits with median readout and CNOT error at the time of experiments. Calibrations can be found at \cite{cal}}.
    \label{fig:manila}
\end{figure}

\begin{figure}[h!]
\begin{minipage}[t]{0.475\columnwidth}
  \includegraphics[width=\linewidth]{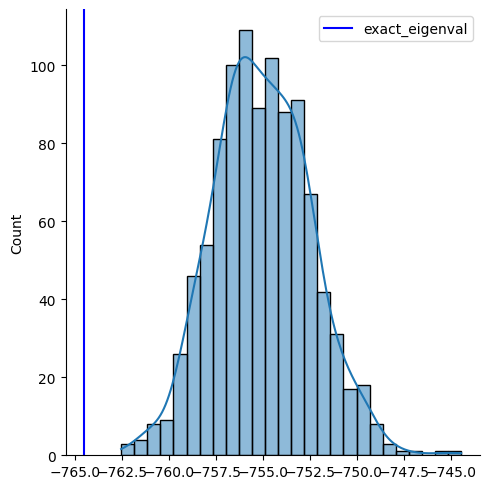}
  \caption{Noisy simulation with 1000 samples.}
  \label{fig:noisy}
\end{minipage}\hfill 
\begin{minipage}[t]{0.475\columnwidth}
  \includegraphics[width=\linewidth]{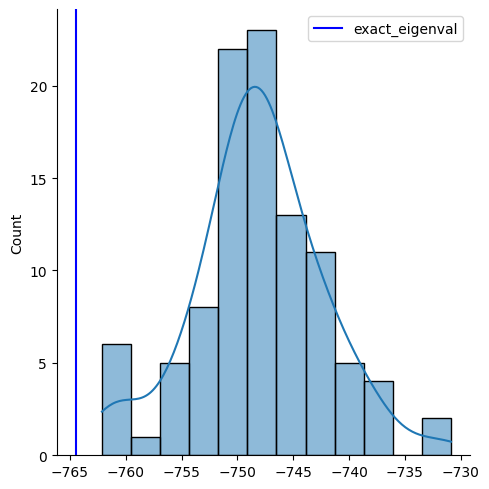}
  \caption{Hardware run with 100 samples.}
  \label{fig:hardware}
\end{minipage}
\end{figure}

To perform noisy simulations, we use \texttt{ibmq manila} with the latest quantum hardware configuration. The quantum system consists of 5 qubits in a row as shown in Fig.~\ref{fig:manila}. A total of 1000 samples were taken on the noisy simulator with M3 error mitigation. For the hardware run, we were only able to take 100 samples only due to high queuing time and used T-REx method for readout error mitigation. Fig.~\ref{fig:noisy}, \ref{fig:hardware} and Table.~\ref{table:res_co2_mn2} shows the results. The exact$\_$eigenval vertical line in the figures is the target energy for the VQE algorithm. 

From the Table.~\ref{table:res_co2_mn2}, we can observe that the relative errors in the noisy simulations are consistently lower across all metrics than the corresponding errors in the hardware runs. This could be attributed to the presence of inherent noise and imperfections in the physical quantum hardware. Additionally, increasing the number of samples might provide a more precise estimation of the relative error for better comparison with the noisy simulations. One can also use more sophisticated error mitigation techniques to further reduce the error. 

\begin{table}[h!]
\centering
\begin{tabular}{||c c c c||} 
 \hline
 Execution Type & Min & Avg & Max \\ [0.5ex] 
 \hline\hline
 Noisy Simulation ($10^{-3}$) & 9.77 & 15.38 & 23.29 \\ 
 \hline
 Hardware ($10^{-3}$) & 10.0 & 20.57 & 33.45  \\ 
 \hline
\end{tabular}
\caption{Relative error obtained on noisy simulation and hardware run for CO$_2$ and Mn(II)}
\label{table:res_co2_mn2}
\end{table}

\subsection{Simulation with Amine groups}
\begin{figure}[h!]
    \centering
    \includegraphics[width=0.7\linewidth]{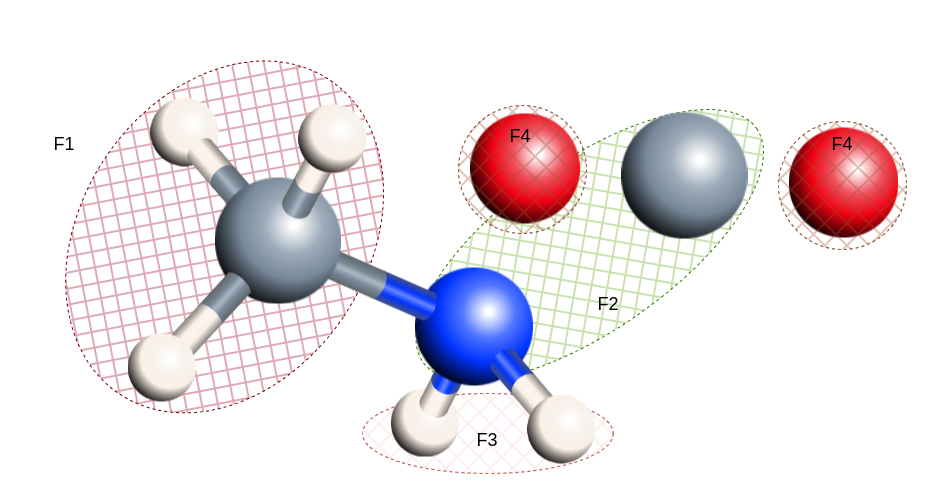}
    \caption{Fragmentation of CH$_3$NH$_2$. F1: methyl (-CH$_3$), F2: NC (Nitrogen and Carbon), F3: HH (two hydrogen atoms bonded to N),  and  F4: OO (two Oxygen atoms of the CO$_2$ molecule). Visualised with \cite{openmx}.}
    \label{fig:amine_frag}
\end{figure}

We performed DMET calculations on CH$_3$NH$_2$ + CO$_2$ system. 4 fragments were created as shown in Fig.~\ref{fig:amine_frag}. STO-3G basis set is used during the calculations. Results are shown in Table.~\ref{table:amine}

\subsubsection{Classical DMET}
Here, each fragment is solved using CCSD as an electronic structure solver. We use Meta-Lowdin as an orbital localization technique. Each iteration optimises the global chemical potential $\mu_{global}$ such that the electrons in the fragments are closer to the actual number of electrons in the system. The optimisation uses the Newton-raphson method. We call this DMET-CCSD.

\subsubsection{Classical DMET with Active Space Reduction}
By considering orbitals near the HOMO-LUMO level, the active space is defined from HOMO-1 to LUMO+1 for the fragment containing Nitrogen and Carbon. Degeneracies were not taken into account. We call this Reduced DMET-CCSD.

\subsubsection{DMET with VQE as fragment solver (Noiseless simulations)}
\begin{figure}[h!]
    \centering
    \includegraphics[width=0.7\linewidth]{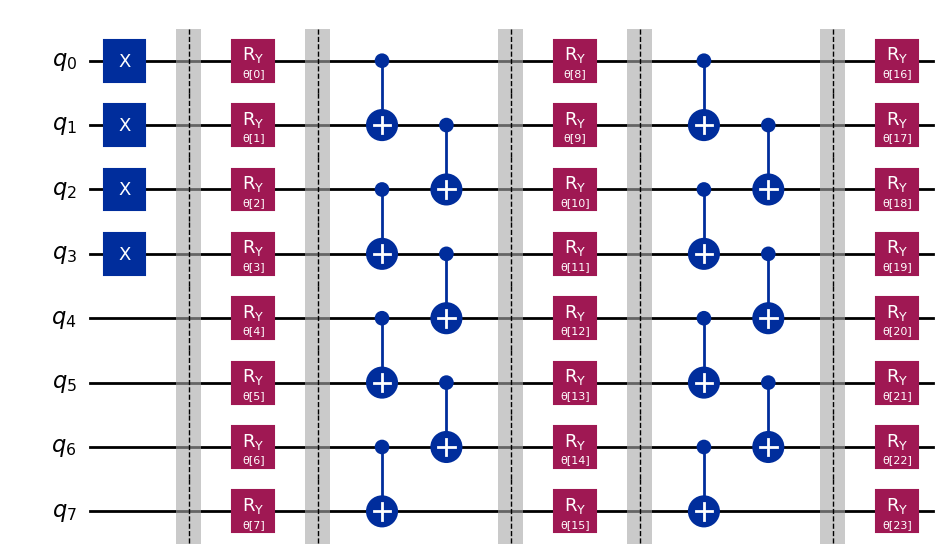}
    \caption{Hardware-efficient ansatz with 2 layers, 8 qubits and initial Hartree-Fock state}
    \label{fig:amine_circ}
\end{figure}

Each fragment is solved using CCSD except for the fragment containing Nitrogen and Carbon which is solved with the VQE algorithm. For this fragment, we consider the active space same as Reduced DMET-CCSD and Jordan Wigner fermion-to-qubit mapping which gives an 8 qubit quantum circuit shown in Fig.~\ref{fig:amine_circ}. We use the L-BFGS-B optimizer for classical optimization. We call this DMET-VQE.

\begin{table}[h!]
\centering
\begin{tabular}{||c c c c||} 
 \hline
 Algorithm & DMET-CCSD & Reduced DMET-CCSD & DMET-VQE \\ [0.5ex] 
 \hline\hline
 Rel Error & $0.92$ & $0.73$ & $1.30$ \\ 
 \hline
\end{tabular}
\caption{Different DMET algorithms for CH$_3$NH$_2$ + CO$_2$ system. Rel Err is in $10^{-3}$ averaged over 10 runs.}
\label{table:amine}
\end{table}

The DMET-CCSD and Reduced DMET-CCSD algorithms achieved lower relative errors indicating better accuracy. On the other hand, DMET-VQE obtained a higher relative error compared to the other two, suggesting a slightly less accurate performance. However, it is worth noting that for the DMET-VQE algorithm, the mean relative error was around $0.79\times10^{-3}$ in the best 8 out of 10 runs, indicating a consistent and reliable performance. The remaining 2 runs resulted in a higher mean relative error of $3.32\times10^{-3}$, which could be attributed to possible variations in the system or the algorithm's sensitivity to specific initial conditions though we should perform more experiments to strengthen the conclusion.

\subsection{DMET for CO$_2$ + Cu-MOF-74}
\begin{figure}[h!]
    \centering
    \includegraphics[width=0.6\linewidth]{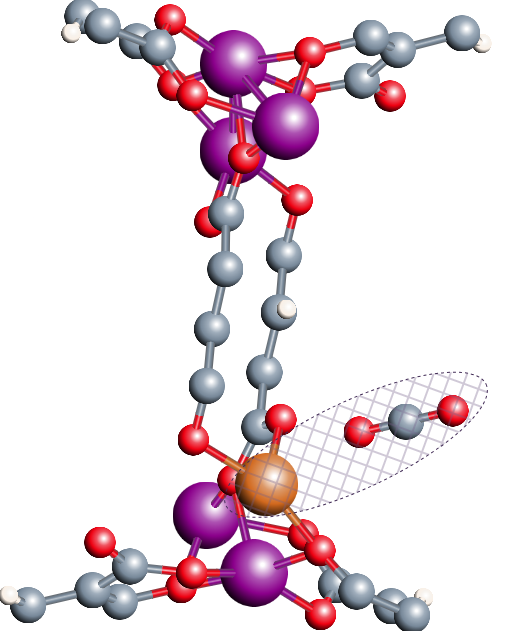}
    \caption{Cu and CO$_2$ are in one fragment (highlighted). A total of 13 fragments. Visualised with \cite{openmx}.}
    \label{fig:mof_frag}
\end{figure}

For scaling the calculation from one binding site to at least one 2D unit cell of the MOF family, we use the DMET algorithm. We explored the DMET method on CO$_2$ + Cu-MOF-74 using a heuristic fragmentation strategy (Fig.~\ref{fig:mof_frag}). Neighbouring atoms were kept in a fragment and the maximum number of atoms in a fragment was 8. We choose the minimal basis set STO-3G due to limited computing power. 
    
We were able to run the algorithm in a feasible time and compared the results with the Reduced-CCSD method (CCSD was performed on the whole molecule with active space reduction from HOMO-1 to LUMO+1). The results are shown in Table.~\ref{table:mof}. To further reduce the computational complexity, the Reduced DMET-CCSD and DMET-VQE perform active space reduction on all the fragments including the binding site. The results demonstrate the feasibility of the DMET approach and show that DMET-VQE outperforms Reduced DMET-CCSD in terms of accuracy. Further analyses and statistical evaluation would be beneficial to solidify these findings. A major bottleneck was computing the fragment Hamiltonians. The runtime was $\sim 50$ minutes suggesting a need for efficient implementation.

\begin{table}[h!]
\centering
\begin{tabular}{||c c c||} 
 \hline
 Algorithm & Reduced DMET-CCSD & DMET-VQE \\ [0.5ex] 
 \hline\hline
 Rel Error ($10^{-3}$) & $0.30$ & $0.16$ \\ 
 \hline
\end{tabular}
\caption{Different DMET algorithms for CO$_2$ + Cu-MOF-74 system.}
\label{table:mof}
\end{table}

\subsubsection{Resource Estimation}

Table.~\ref{table:res_est} presents the estimated resources varying the HOMO- to LUMO+ levels. As the subset from HOMO- to LUMO+ orbitals increases, the number of Hamiltonian terms increases exponentially. The significant rise in Hamiltonian terms from 361 to 14243 suggests that larger active spaces demand more quantum resources. Considering the trade-off between circuit width and Hamiltonian terms, we experimented with only 8 qubit circuit.

\begin{table}[h!]
\centering
\begin{tabular}{||c c c c c||} 
 \hline
 HOMO-LUMO & 1 & 2 & 3 & 4\\ [0.5ex] 
 \hline\hline
 Circuit Width & 8 & 12 & 16 & 20 \\ \hline
 Hamiltonian Terms & 361 & 1819 & 5789 & 14243\\
 \hline
\end{tabular}
\caption{Resource estimation for different active spaces.}
\label{table:res_est}
\end{table}

\section{Conclusion}
The research uses quantum computing techniques for studying carbon capture on Metal-Organic Frameworks by investigating the minimum of the Potential Energy Surface (PES) of MOFs and CO$_2$. We used Hardware-efficient ansatz for the VQE algorithm with a deparameterisation approach to freeze $R_y$ gates with standardized parameter values which helped in simplifying the energy landscape and reducing the trainable parameters while maintaining the accuracy of the global minimum. DMET was used to simulate complex molecules like Amines and MOF unit cells classically as well as in quantum using VQE as a fragment solver. Additionally, we present findings from noisy simulations and hardware experiments with the M3 and T-REx error mitigation schemes. We also performed a resource estimation when higher HOMO-LUMO levels are taken into consideration. Overall, this research paves the way for leveraging quantum computing in the design of complex materials, specifically for practical applications such as carbon capture. In future, we intend to conduct more robust experiments with efficient code implementation and apply the techniques used here to investigate problem-inspired circuit ansatzes and compare them with hardware-efficient ansatzes in subsequent work.

\section*{Acknowledgment}
We acknowledge the hardware support from IBM Quantum provided. Quantistry \cite{quantistry} was used to obtain reference classical PES and optimized geometry of molecules. Qiskit \cite{Qiskit} and Tangelo \cite{tangelo} were used to perform quantum simulations. We also want to thank Xanadu and Deloitte for organising QHack 23 and Quantum Climate Challenge 23 respectively creating an environment that facilitated the progress of this work.

\bibliographystyle{IEEEtran}
\bibliography{references}

\vspace{12pt}

\end{document}